\documentstyle[12pt,epsfig]{article}
\newcommand{\subs}[1]{{\mbox{\scriptsize #1}}}
\begin{document}

 \newcommand \be {\begin{equation}}
\newcommand \ee {\end{equation}}
 \newcommand \ba {\begin{eqnarray}}
\newcommand \ea {\end{eqnarray}}

\newcommand{\siml}{\stackrel{\langle}{\sim}}

\title{Back to basics: historical option pricing revisited}

\author{Jean-Philippe Bouchaud$^{1,2}$, Marc Potters$^2$}
\maketitle

\begin{center}
\begin{tabular}{ll}
$^1$ & Service de Physique de l'\'Etat Condens\'e, \\
& Centre d'\'etudes de Saclay, Orme des Merisiers, \\
& 91191 Gif-sur-Yvette C\'edex, {\sc france} \\
& E-mail: bouchau@amoco.saclay.cea.fr \\
$^2$ & Science \& Finance, 109-111 rue Victor Hugo,\\
& 92523 Levallois Cedex, {\sc france}\\
& E-mail: marc.potters@science-finance.fr
\end{tabular}
\end{center}

\begin{abstract}

We reconsider the problem of option pricing using historical probability
distributions. We first discuss how the risk-minimisation scheme proposed
recently is an adequate starting point under the realistic assumption that
price increments are uncorrelated (but not necessarily independent) and
of arbitrary probability density. We discuss in particular how, in the Gaussian limit, the Black-Scholes results are recovered, including the fact that the
average return of the underlying stock disappears from the price (and the hedging strategy). We compare this theory to real option prices and find these reflect in a surprisingly accurate way the subtle
statistical features of the underlying asset fluctuations.

\end{abstract}
\newpage

\section{Introduction}

The famous Black and Scholes option pricing theory has two remarkable
features: the hedging strategy eliminates risk entirely, and
the option price does not depend at all on the average return of the
underlying
asset \cite{BSc,Hull,Wilmott}. The second property means that the option price
is not simply the actualized 
average of the future pay-off over the historical probability distribution,
which 
obviously would depend on the average return. This is even more striking in
the case of the
Cox-Ross-Rubinstein binomial model \cite{CRR,Hull} where the pricing measure is
completely unrelated to
the actual distribution of returns. This has lead to a rather abstract and
general framework
for derivative pricing, where the absence of arbitrage opportunities leads,
for models
where risk can be eliminated completely, to the existence of a `risk-neutral
probability measure' 
(unrelated to the historical one) over which the relevant average should be
taken to obtain the price of derivatives \cite{Pliska,Baxter}. It is thus a
rather common belief that the knowledge of the `true' 
probability distribution of returns is a useless information to price options.
The credence is rather that the relevant `implied' value of the parameters
should be obtained from option market themselves, and used to price other
instruments (for example exotic options) \cite{Hull,Dupire}.  

However, in most models of stock fluctuations, except for very special cases
(continuous time Brownian motion and binomial, both being very poor
representation of the reality), risk in option trading cannot be eliminated,
and strict arbitrage opportunities do not exist, whatever the price of the
option. That risk cannot be eliminated is furthermore the fundamental reason
for the very existence of option markets. It would thus be more satisfactory
to have a theory
of options where the true historical behaviour of the underlying asset was used
to compute the
option price, the hedging strategy, and {\it the residual risk}. The latter is
clearly important to estimate, both for risk control purposes, but also
because it is reasonable to think that this residual risk partly determines
the bid-ask spread imposed by market-makers.
The natural framework for this is the risk minimisation approach developed by
several authors \cite{Schw,Schal,BS,BP,AS,WO,Matacz}, where the optimal trading
strategy is determined such that the chosen measure of risk (for
example the variance of the wealth balance) is minimized. The
`theoretical' price is then obtained using a fair game argument. Note that in this
approach, the option price is not unique since it depends on the definition of
risk; furthermore, a risk-premium 
correction to the fair game price can be expected in general. From a
theoretical point of view, this would generally be regarded as a lethal
inconsistency. From a practical point of view, however, we see this as an
advantage: since the price ambiguity is a constitutive property of option
markets, it is interesting to understand the origin and size of this
ambiguity. In this framework, the historical probability distribution
determines the 
`pricing kernel' to be used in the option price formula. We show in detail
how, in the Black-Scholes limit, the average trend indeed completely
disappears from the formula, and all
the classical results are recovered. For more general models, however, the
independence of the price on the average return is non trivial.

The outline of this paper is as follows. We first recall the basic steps
leading to option
prices and optimal hedges for a general process with uncorrelated (but
not necessarily independent) increments,
which we present in terms of a cumulant expansion to show how the Black-
Scholes results are obtained in the 
corresponding Gaussian limit. The first cumulant correction provides a theory
for the volatility smile in terms of the (maturity dependent) kurtosis of the
terminal price distribution. We compare this theory to real option prices (on
a liquid market) and find that 
these option prices reflect in a surprisingly accurate way the subtle
statistical features 
of the underlying asset fluctuations \cite{Olsen,review}, in particular the
persistent nature of the volatility fluctuations.

\section{A risk minimisation theory of option pricing}

\subsection{The global wealth balance}
\label{s:wealth-balance}

Let us first write the wealth balance equation corresponding to the writing of
a European call option.
At time $t=0$, the writer receives the
price of the option ${\cal C}[x_0,x_s,T]$, on a
certain asset which value is
$x(t=0)=x_0$. The strike price is $x_s$. Between $t=0$ and $t=T$,
the writer trades the underlying asset at discrete times
$t=k\tau, \ k=1,...,N=T/\tau$; his strategy is to hold
$\phi_k(x_k)$ assets if the price is $x(t)=x_k$ when the time is $t=k\tau$.
It is easy to show that the change of wealth due to this trading is given by
\cite{BP}:
\be
\Delta W_\subs{trading} = \sum_{k=0}^{N-1} \phi_k(x_k) [x_{k+1}-e^{r\tau}x_k]
e^{r(T-t_k-\tau)}
\label{deltaWtrading}
\ee
where
$r$ is risk-free  rate and $t_k=k \tau$. At time $T=N\tau$, the writer looses the
difference $x_N-x_s$ if the option is exercized. Thus the complete wealth
balance
reads:
\be
\Delta W =  {\cal C}[x_0,x_s,T] e^{rT} - \max(x_N-x_s,0) + \sum_{k=0}^{N-1}
\phi_k(x_k)  \delta x_ k e^{r(T-t_{k+1})}
\label{deltaWtotal}
\ee
where we have introduced the notation: $\delta x_k \equiv
[x_{k+1}-e^{r\tau}x_k] $.
 Note that $\delta x_k$ is posterior to the instant $k$ where $\phi_k$ is
determined.
Denoting as $\langle ... \rangle$ the average over the historical
distribution, the average profit is given by:
\be
 \langle\Delta W\rangle =  {\cal C}[x_0,x_s,T] e^{rT} - \langle\max(x_N-x_s,0)\rangle + 
\sum_{k=0}^{N-1}
\langle\phi_k(x_k)\rangle \langle \delta x_k \rangle e^{r(T-t_{k+1})}
\label{profit}
\ee
The fair game requirement then fixes ${\cal C}[x_0,x_s,T]$ such that $\langle\Delta
W\rangle=0$.

Although other interesting definitions could be considered \cite{BP}, we
restrict here
to the case where the risk is measured as:
\be
R \equiv \langle\Delta W^2\rangle - \langle\Delta W\rangle^2 = \langle\Delta W^2\rangle
\label{risk}
\ee
The risk  $R$ is always greater than or equal to zero and the minimum is
obtained
for a certain optimal strategy $\phi^*$, determined by a functional derivation
of (\ref{risk}) with respect to 
with respect to $\phi(x,t)$. This determines the option price through:
\be
 {\cal C}[x_0,x_s,T]  = e^{-rT} \langle\max(x_N-x_s,0)\rangle - \sum_{k=0}^{N-1}
\langle\phi_k^*(x_k)\rangle \langle \delta x_k \rangle e^{-rt_{k+1}}
\label{price}
\ee
Note that since $\phi^*$ depends a priori on our choice of the variance as the
relevant 
measure of risk, the price of the option is not unique, but reflects (among
other things) the operator's perception of risk.

\subsection{The case of zero excess average return}
\label{s:mu-zero}
In this section we consider the case where the average return
of the stock over the bond, $m=\langle \delta x_k \rangle$, is
zero. This simplifying hypothesis is often justified in practice for
small maturities, where average return effects are small compared to
volatilites, and can be treated perturbatively, as shown in section~
\ref{s:mu-small}.

Let $P(x,T|x_0,0) dx$ be the probability that the asset value is $x$ at time
$T$, knowing that it was $x_0$ at time $0$. When $m=0$, Eq. (\ref{price}) then
yields an option price independent of the trading strategy:
\be 
{\cal C}[x_0,x_s,T;m=0] = e^{-rT} \langle \max(x_N-x_s,0) \rangle \equiv
e^{-rT} \int_{x_s}^\infty dx (x-x_s) P(x,T|x_0,0)
\label{pricemequal0}
\ee
Note that our assumption that $m=0$ means that the average of $P(x,T|x_0,0)$
is
not at $x_0$, but at the forward price $x_0 e^{rT}$.

In order to proceed with the risk-minimization,
we shall assume that the price
increments $\delta x_k$ are uncorrelated random variables, such that
$\langle \delta x_k \delta x_\ell \rangle = \sigma^2 \delta_{k,\ell}$,
where $\delta_{k,\ell}$ is the Kronecker symbol. Assuming that $\sigma$ does not depend on $k$ is in
general not justified, since it amounts to assuming that share
price follows an additive random process of constant volatility (but
with an arbitrary distribution for the increments). 
Actually, real data is often closer (for short maturities) to being an 
additive random process rather than a multiplicative one \cite{BP}, an
assumption which does introduce an spurious positive skew in the price
distribution. In reality, however, $\sigma$ depends on $k$, which reflects
{\sc arch}-like effects (or time
persistent volatility \cite{Olsen}). Taking this effect into account would lead to more involved calculations, which can
however still be completed analytically \cite{BP}. With these approximations
in mind, the relevant formula for risk is rather simple:
\ba\nonumber
\langle \Delta W^2 \rangle &=& \langle \Delta W^2 \rangle_0 + \sigma^2 \sum_{k=0}^{N-1}
\int_{0}^\infty dx P(x,t_k |x_0,0) \phi_k^2(x) e^{2r(T-t_{k+1})} \\ \nonumber
&-& 2 e^{r(T-t_{k+1})} \sum_{k=0}^{N-1}  \int_{0}^\infty dx P(x,t_k
|x_0,0) \phi_k(x) \\
& &\int_{x_s}^{\infty} dx' (x'-x_s) P(x',T|x,t_k) \langle
\delta x_k \rangle_{x,t_k \to x',T} 
\label{risk0}
\ea
where $\langle \Delta W^2 \rangle_0$ is the unhedged ($\phi_k \equiv 0$) risk
associated to the option, and $\langle \delta x_k \rangle_{x,t_k \to x',T}$ is
the conditional average of $\delta x_k$, on the trajectories starting 
at $x$ at time $t_k$ and ending at point $x'$ at time $T$.

The optimal trading strategy is obtained by setting \cite{BS,BP,AS}:
\be
\frac{\partial \langle \Delta W^2 \rangle}{\partial \phi_k(x)} = 0
\ee
for all $k$ and $x$. This leads to the following explicit result for the
optimal
hedging strategy:
\be
\phi_k^*(x)= \frac{e^{-r(T-t_{k+1})}}{\sigma^2} \int_{x_s}^{\infty} dx' (x'-x_s)
\langle \delta x_k \rangle_{x,t_k \to x',T} P(x',T|x,t_k)
\label{optimalstrategy0}
\ee
This formula simplifies somewhat when the increments are Gaussian, and one
finally 
finds the famous Black-Scholes `$\Delta-$hedge': $\phi_k^*(x)=\partial {\cal
C}
[x,x_s,T-t_k]/\partial x$. In the non-Gaussian case, however, this simple
relation between the derivative of the option price and the trading strategy
no longer holds (see Eq. (\ref{strat2}) below).

Inserting (\ref{optimalstrategy0}) into (\ref{risk0}) leads to the following
formula for the residual risk:
\be
R^* = \langle \Delta W^2 \rangle_0 - D\tau
\sum_{k=0}^{N-1}
\int_{0}^\infty dx P(x,t_k|x_0,0) \phi_k^{*2}(x)
e^{-r(T-t_{k+1})}
\label{residualrisk0}
\ee
In general, the left-hand side of
(\ref{residualrisk0}) is non-zero; in practice it is even quite high -- for
example, for typical one-month options on liquid markets, $\sqrt{R*}$ represents as much as $25\%$ of the option price itself \cite{BP}.
However, in the special case where $P(x,t|x_0,0)$ is normal (or log-normal),
and in
the limit of continuous trading, that is, when $\tau \to 0$, one can show that
the residual risk $R^*$ actually vanishes, thanks 
to a somewhat miraculous identity for Gaussian integrals \cite{BS}. Hence, the
above formalism matches smoothly with all the Black-Scholes results in the
limit of a continuous time
Brownian (or log-Brownian) process, at least when the excess average return of
the
asset is zero. Let us now discuss how these results are changed if the average
return $m \equiv \langle \delta x_k
\rangle$ is non zero (but small).

\subsection{Small non-zero average return}
\label{s:mu-small}

More precisely, we shall consider the case where $m N \ll \sigma \sqrt{N}$
($N=T/\tau$),
or, more intuitively, that the average return on the time scale of the option
is  small compared to the typical variations, which is certainly the case for
options up to a few
months{\footnote{Typically, $m=5
\%$ annual and $\sigma=15
\%$ annual. The order of magnitude of the error made in neglecting the
second
 order term in $m$ is 
$m^2 N/\sigma^2
\simeq 0.1$ even for $N \tau=1$ year.}}. The global wealth balance then
includes the  term related to the trading strategy, which reads\footnote{For
the sake of simplicity, we shall set the interest $r$ to zero in the
following. See \cite{AS} for a more complete discussion.}:
\be
\langle \Delta W_\subs{trading} \rangle  = m \sum_{k=0}^{N-1} \int_0^\infty dx
P(x,t=k\tau |x_0,0) \phi_k^*(x) 
\label{deltw}
\ee
The advantage of considering a small average return is that one can
do a perturbation around the zero
average return case, and still use the explicit optimal strategy of
(\ref{optimalstrategy0}) to lowest order in $m$.

Compared to the case $m=0$, the option price is changed both because
$P(x,k|x_0,0)$ is {\it biased}, and
 because $\langle \Delta W_\subs{trading} \rangle$
must be substracted off from Eq. (\ref{price}).
It is convenient to use the Fourier transform of the
probability distribution $\tilde{P}(z)=\int_{-\infty}^{\infty}
dx P(x,N|x_0,0) \exp(i x z)$ and to
expand it in a series introducing
 the {\it cumulants} $c_n$. They are defined by
\be
  \tilde{P}(z)=\exp\Bigr[\sum_{n=1}^{\infty}\frac{c_n (iz)^n}{n!}
\Bigl],
 \label{cumul}
\ee
where $c_1=mN$, $c_2=N\sigma^2$, $\kappa=c_4/c_2^2$ is
 the kurtosis, etc... Applying the cumulant expansion to
 the probability distribution in Eq.(\ref{optimalstrategy0}),
 we obtain the following optimal strategy for $m=0$ \cite{BP}:
\be
 \phi_k^*(x)  = \frac{1}{c_2}
\sum_{n=2}^\infty
\frac{(-1)^n c_n}{ (n-1)!} \frac{\partial^{n-1}}{\partial
x^{n-1}} {\cal C}[x,x_s,T-t_k]
\label{strat2}.
\ee
Note again that in the Gaussian case ($c_n=0$ for $n>2$), one recovers the 
standard Black and Scholes `$\Delta-$hedge'.

Inserting the optimal strategy (\ref{strat2}) into Eq.(\ref{deltw}) and
integrating by parts one gets an expansion of the trading term
$\langle \Delta W_\subs{trading} \rangle$. Inserted into
Eq.(\ref{price}), this gives the following
correction to the option price \cite{BP}:
\begin{eqnarray}
{\cal C}[x_0,x_s,T;m] &=& {\cal C}[x_0,x_s,T;m=0]\\ \nonumber
&& - \frac{m}{c_2} \sum_{n=3}^\infty
\frac{c_n}{ (n-1)!} \frac{\partial^{n-3}}{\partial
x'^{n-3}}P_0(x',N|x_0,0)|_{x'=x_s} +{\cal O}(m^2)
\label{m}
\end{eqnarray}
where the shorthand $P_0$ stands for the probability
distribution where the first cumulant $m$ has been set to zero.

In the Gaussian case, $c_n=0$ for all $n \geq 3$,
and one thus sees
explicitly
 that ${\cal C}_m = {\cal C}_0$, at least to first
order in $m$. Actually, one can show that this is true to all orders in $m$ in
the
Gaussian case, which is an alternative way to derive the result of Black and
Scholes in a Gaussian context \cite{BP} {\footnote{This result is
obvious on the framework of Ito's calculus, which is only valid
for all Gaussian processes (including the log-normal) in the continuum time limit.}}.

However, for even distributions with fat tails ($c_3=0$ and $c_4>0$), it is
easy to see from the above formula that a positive average return $m > 0$
increases the price of out-of-the-money options
 ($x_s > x_0$), and decreases the
price of in-the-money options ($x_s < x_0$).
 Hence, we see again
explicitly
  that the independence of the option price on the
 average return $m$, which is one of the
most important result of Black and Scholes,
 does not survive for more general
models of stock fluctuations.

Note finally that Eq. (\ref{m}) can also be written as:
\be
 {\cal C}[x_0,x_s,T;m] =  
  \int_{x_s}^\infty dx (x-x_s) Q(x,T|x_0,0)
\label{BS_price_analogy}
\ee
with an effective distribution $Q$ defined as:
\be
Q(x,T|x_0,0) = P_0(x,T|x_0,0) -  \frac{m}{c_2}
\sum_{n=3}^\infty
\frac{c_n}{ (n-1)!} \frac{\partial^{n-1}}{\partial
x^{n-1}}P_0(x,N|x_0,0)
\label{effectivedistribution}
\ee
The integral over $x$ of $Q$ is one, but $Q$ is
not a priori positive everywhere. This means that for a certain family of pay-
offs, the fair price of the option (\ref{BS_price_analogy}) may be negative in
the absence of risk-premium. From a practical point of view, however, this
requires rather absurd values for the average return and for the strike price,
which in turn would lead to a large residual risk. 

The pseudo-distribution  $Q$ generalizes the `risk neutral probability' 
usually discussed
in the context of the Black-Scholes
theory, and also has the property that the excess average return 
(the integral of $(x-x_0) Q$ over $x$) is zero, as
can easily be seen by inspection from (\ref{effectivedistribution}).
In fact, one can derive a general formula for $Q$ without any restriction
on $m$ of $r$ \cite{AS,WO}, and the effective distribution still has the properties
that it is normalized to one and has zero average excess return.

\section{Volatility smile and implied kurtosis}
\label{s:empirical}

\begin{figure}
\centerline{\hbox{\epsfig{figure=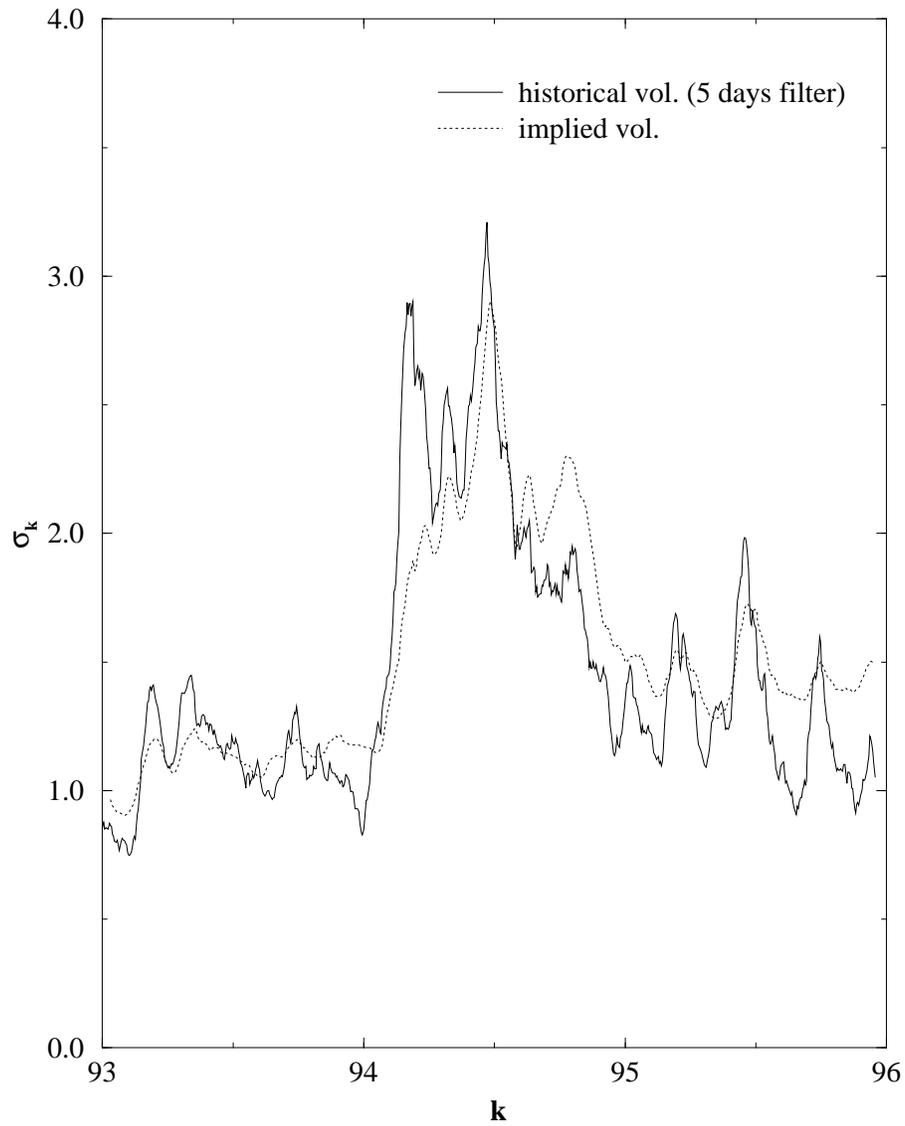,width=10cm}}}
\vskip 0.5cm
\caption{\small{Comparison between the historical volatility of the {\sc bund}
(measured from high frequency data and filtered over the past five days), and
the implied volatility, extracted from the option prices through formula
(\protect\ref{smile}).}}
\label{fig1}
\end{figure}

In the case where the market fluctuations are moderately non-Gaussian, as is
the case for liquid markets, one
might expect that the first terms in the cumulant expansion around the Black-Scholes formula are sufficient to account for real option prices. If one only
retains the leading order correction which is (for symmetric fluctuations)
proportional to the kurtosis $\kappa$, one finds that the 
price of options ${\cal C}(x_0,x_s,T)$ can be written as a Gaussian
Black-Scholes
formula\footnote{Note that the operators rather use the more standard log-normal Black-Scholes
formula, which, as noted above, induces a spurious positive skew not present
in
real data (at least for short maturities). In order to correct for this skew,
the
log-normal volatility smile is then negatively skewed. A more symmetric smile is observed
if 
one talks in terms of a Gaussian volatility, which is what we adopt in the following.}, but with a modified value of the volatility $\sigma$, which
becomes price and maturity dependent \cite{Potters}:
\be 
\sigma_\subs{imp}(x_s,T) = \sigma
\left[1+ \frac{\kappa_T 
}{24}
\left(\frac{(x_s-x_0)^2}{\sigma^2 T}- 1
\right)\right] 
\label{smile}
\ee 
The volatility $\sigma_\subs{imp}$ is called the implied volatility by the market
operators, who use the standard Black-Scholes formula to price options, but
with a
value of the volatility which they estimate intuitively, and which turns out
to depend on
the exercice price in a roughly parabolic manner, as indeed suggested by Eq.
(\ref{smile}). This is the famous
`volatility smile'. Eq. (\ref{smile}) furthermore shows that the curvature of
the smile is directly related to the
kurtosis $\kappa_T$ of the underlying statistical process on the scale of the maturity
$T=N\tau$. We have tested this prediction by directly comparing the `implied
kurtosis' \cite{Corrado}, obtained by extracting from real option prices the volatility
$\sigma$ and the curvature of the
implied volatility smile, to the historical value of the volatility and of the
kurtosis $\kappa_N$. We have mostly studied short maturities (up to two
months) options on futures, for which the interest rate can be set to zero. We
also restrict to liquid markets (such as the {\sc bund} option market) where
(i) non Gaussian effects are not too strong, and (ii) risk-premiums are expected to be small, and thus where a comparison with the fair price is
meaningful. 

We have found the following results. The implied volatility turns out to be
highly correlated with a short time filter of the historical volatility: see
Fig. 1. Fig. 2 shows the comparison between the 
implied and historical kurtosis, {\it with no further ajustable parameters}. Note
that the historical kurtosis decays more slowly than $N^{-1}$, which would be
expected for a process with independent, identically distributed increments.
This anomalously slow decay is directly related to volatility persistence
effects \cite{Potters,BP}.

\begin{figure}
\centerline{\hbox{\epsfig{figure=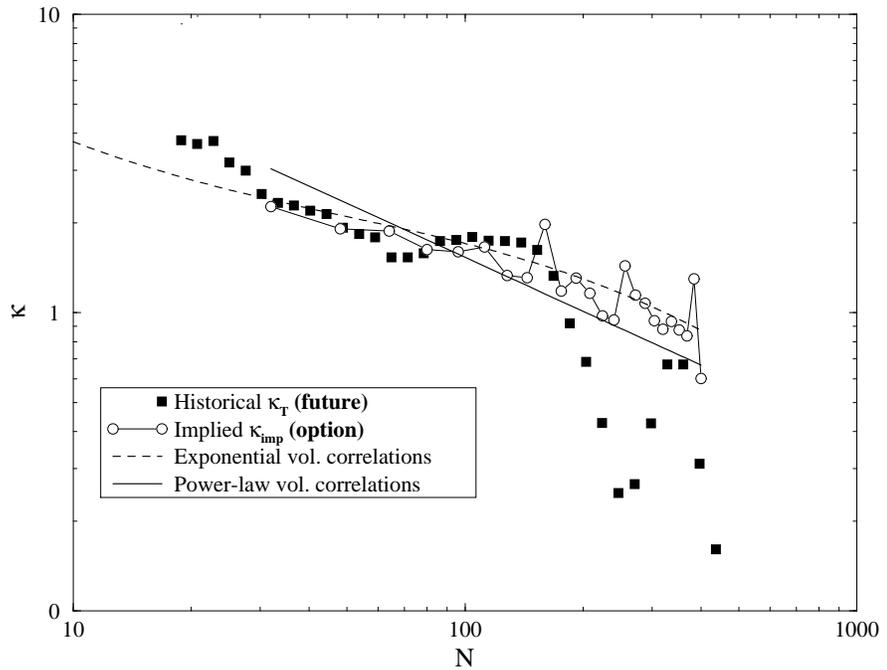,width=10cm,angle=270}}}
\vskip 0.3cm
\caption{\small{Plot (in log-log coordinates) of the average implied kurtosis
$\kappa_\subs{imp}$ (determined by fitting the implied volatility for a fixed
maturity by a parabola)
and of the empirical kurtosis $\kappa_N$ (determined directly
from the
historical movements of the {\sc bund} contract), as a function of the
reduced time
scale $N=T/\tau$, $\tau = 30$ minutes. All
transactions of options on the
{\sc bund} future from 1993 to 1995 were analyzed
along with 5 minute tick data of the
{\sc bund} future for the same period. We show for comparison a fit with 
$\kappa_N \simeq N^{-0.6}$ (dark line). A fit with an exponentially decaying
volatility correlation function is however also acceptable (dotted line).
}}
\label{fig3}
\end{figure}

It is interesting to note that the kurtosis correction to the optimal strategy
does not coincide with the market practice of using the implied volatility in the Black-Scholes $\Delta-$hedge. However, since the risk is minimum for $\phi=\phi^*$, this means that the increase of risk due to a small error $\delta \phi$ in the strategy is only of order $\delta \phi^2$, and thus often quite
small in practical applications.

The remarkable agreement between the implied and historical
value of the parameters (which we have also found on a variety of other assets),
and the fact that they evolve similarly with maturity, shows 
that the market as a whole is able to correct (by trial and errors) the
inadequacies of the Black-Scholes formula, and to encode in a satisfactory way
both the fact that the distribution has a positive kurtosis, and that this
kurtosis decays with maturity in an anomalous fashion due to volatility persistence effects.

\section{Conclusion}

In our opinion, mathematical finance in the past decades has overfocused
on the concept of arbitrage free pricing, which relies on very specific models
(or instruments) where risk can be eliminated completely. This leads to a
remarkably elegant and consistent formalism, where derivative
pricing amounts to determining the risk-neutral probability measure, which in
general does not coincide with the historical measure. In doing so, however,
many important and subtle features are swept under the rug, in particular the
amplitude of the residual risk. Furthermore, the fact that the
risk-neutral and historical probabilities need not be the same is often 
an excuse for not worrying when the parameters of a specific model deduced
from derivative markets are very different from historical ones. This is
particularly obvious in the case of interest rates \cite{IRC}. 
In our mind, this rather reflects that an important effect has been left out
of the models, which in the case of interest rates is a risk premium effect \cite{IRC}. We
believe that a more versatile (although less elegant from a mathematical point
of view) theory of derivative pricing, such as the one discussed above, 
allows one to use in a consistent and fruitful way the empirical data
on the underlying asset to price, hedge, and control the risk the
corresponding 
derivative security. Extension of these ideas to interest rate derivatives is
underway.

\section{Acknowledgements}
We thank J.-P. Aguilar, E. Aurell, P. Cizeau, R. Cont, L. Laloux and J. Miller
for many important discussions. We also thank P. Wilmott for giving us the
opportunity of writing this paper.

\end{document}